\documentclass[conference]{IEEEtran}
\IEEEoverridecommandlockouts

\usepackage{cite}
\usepackage{amsmath,amssymb,amsfonts}
\usepackage{algorithmic}
\usepackage[pdftex]{graphicx}
\usepackage{textcomp}
\usepackage{xcolor}
\usepackage{multirow}
\usepackage[colorlinks=true, allcolors=blue]{hyperref}
\usepackage{tikz-uml}
\usetikzlibrary{calc}
\usepackage{url}
\usepackage[capitalize]{cleveref}
\usepackage{epstopdf}

\DeclareGraphicsExtensions{.pdf,.png,.jpg}

\DeclareRobustCommand{\IEEEauthorrefmark}[1]{\smash{\textsuperscript{\footnotesize #1}}}
\PassOptionsToPackage{hyphens}{url}\usepackage{hyperref}

\begin{document}

\makeatletter
\newcommand{\linebreakand}{%
      \end{@IEEEauthorhalign}
      \hfill\mbox{}\par
      \mbox{}\hfill\begin{@IEEEauthorhalign}
    }
\makeatother

\title{BenchQC - Scalable and modular benchmarking of industrial quantum computing applications}

\author{
    \IEEEauthorblockN{Florian Geissler\IEEEauthorrefmark{1}, Eric Stopfer\IEEEauthorrefmark{2}, 
    Christian Ufrecht\IEEEauthorrefmark{2}, Nico Meyer\IEEEauthorrefmark{2},
    Daniel D. Scherer\IEEEauthorrefmark{2}, \\
    Friedrich Wagner\IEEEauthorrefmark{2}, Johannes M. Oberreuter\IEEEauthorrefmark{3}, 
    Zao Chen\IEEEauthorrefmark{3}, Alessandro Farace\IEEEauthorrefmark{3},
    Daria Gutina\IEEEauthorrefmark{4},\\
    Ulrich Schwenk\IEEEauthorrefmark{4}, Kimberly Lange\IEEEauthorrefmark{4},
    Vanessa Junk\IEEEauthorrefmark{4}, 
    Thomas Husslein\IEEEauthorrefmark{4}, Marvin Erdmann\IEEEauthorrefmark{5}, \\
    Florian Kiwit\IEEEauthorrefmark{5},
    Benjamin Decker\IEEEauthorrefmark{5}, 
    Greshma Shaji\IEEEauthorrefmark{5}, Etienne Granet\IEEEauthorrefmark{6}, Henrik Dreyer\IEEEauthorrefmark{6},\\
    Theodora-Augustina Dr$\breve{\text{a}}$gan\IEEEauthorrefmark{1},
    and Jeanette Miriam Lorenz\IEEEauthorrefmark{1}}\\
    \IEEEauthorblockA{\IEEEauthorrefmark{1}\textit{Fraunhofer Institute for Cognitive Systems IKS},
     Munich, Germany\\ (Email: florian.geissler@iks.fraunhofer.de)}
     \IEEEauthorblockA{\IEEEauthorrefmark{2}\textit{Fraunhofer Institute for Integrated Circuits IIS},
     Nuremberg, Germany}
    \IEEEauthorblockA{\IEEEauthorrefmark{3}\textit{Reply SE, Machine Learning Reply GmbH},
     Munich, Germany}
    \IEEEauthorblockA{\IEEEauthorrefmark{4}\textit{OptWare GmbH},
    Regensburg, Germany}
    \IEEEauthorblockA{\IEEEauthorrefmark{5}\textit{BMW Group},
    Munich, Germany}
    \IEEEauthorblockA{\IEEEauthorrefmark{6}\textit{Quantinuum},
    Munich, Germany}
    }

\maketitle

\begin{abstract}
We present \textit{BenchQC}, a research project funded by the state of Bavaria,
which promotes an application-centric perspective for benchmarking real-world quantum applications. 
Diverse use cases from industry consortium members are the starting point of a benchmarking workflow, that builds on the open-source platform QUARK, encompassing the full quantum software stack from the hardware provider interface to the application layer.
By identifying and evaluating key metrics across the entire pipeline, we aim to uncover meaningful trends, provide systematic guidance on quantum utility, and distinguish promising research directions from less viable approaches.
Ultimately, this initiative contributes to the broader effort of establishing reliable benchmarking standards that drive the transition from experimental demonstrations to practical quantum advantage.
\end{abstract}



\bstctlcite{IEEEexample:BSTcontrol}
\section{Introduction}
\subsection{Motivation}
Quantum computing (QC) offers new avenues to tackle complex data analysis and optimization problems.
In light of the ever-growing demand for computational resources, this puts the technology at the center of interest for both scientific researchers and industrial end users. Drug discovery, logistics, and aerodynamic design are just a few examples of sectors where quantum technology promises disruptive transformations.
Quantum computing has the potential to boost industrial applications in particular for optimization, simulation, and machine learning~\cite{bayerstadler_industry_2021, riofrio_quantum_2025}. This is realized by solving problems that are currently intractable for classical systems, or by building more efficient solutions for already tractable problems, using for example less learning data.
The Boston Consulting Group estimates that, at tech maturity, quantum computing will unlock hundreds of use cases, with potential value creation of $\$100$B and more in \textit{each} of the above fields~\cite{jean-francois_bobier_what_2021}.
Being at the forefront of this emerging technology is therefore a competitive advantage that elicits strategic investments across various industrial domains. 

With limited quantum hardware available in the current Noisy Intermediate-Scale Quantum (NISQ) era~\cite{preskill_nisq}, quantum utility~\cite{kim_evidence_2023} remains to be demonstrated in real-world applications~\cite{brooks_beyond_2019, lorenz_quest_2024}. As an additional complication, the notion of utility and application realism in itself is widely debated. For practical purposes, the authors of~\cite{herrmann_quantum_2023} propose for example \textit{application readiness levels} to evaluate quantum applications in the pursuit of quantum utility. From an end user perspective, utility ultimately depends on the specific use case scenario, in which a quantum solution delivers a benefit in terms of compute or solution quality.
Benchmarking~\cite{jain1991art} is crucial in this process, providing insight into the practical applicability of quantum solutions. It allows industrial players to track performance improvements, evaluate the scalability of quantum approaches, and ensure that frameworks catch up with the progress of the technology. This is particularly important for companies dealing with complex and large-scale use cases.
A strong benchmarking framework guides investment decisions and sets realistic expectations.
Early involvement in benchmarking initiatives will help industry players to capitalize on advances in quantum technology as the systems become more mature.

In these efforts, it is essential to align on benchmarking procedures that represent realistic problem instances, and span the entire quantum software stack from the hardware interface to the application layer. 
To this day, the international quantum communities have each prioritized individual metrics, concepts, and applications for benchmarking, see Section~\ref{sec:related_work}.
Nevertheless, the informative power remains limited for industrial applications, in particular in emerging fields such as quantum machine learning (QML)~\cite{schreiber_classical_2023, bowles_better_2024}. 
For end users, it can be challenging to understand whether their problem category is well-suited to benefit from future quantum technology, how large the problem can be scaled, or which hardware and algorithms are the best fit.
BenchQC therefore takes a different approach, and makes industrial applications the starting point of the benchmarking workflow. 
Problem instances studied in the BenchQC project range hereby from the optimal distribution and configuration of LiDAR sensors within production plants, over 6G beam management, to the simulation of dynamic fermions in material science (see Sec.~\ref{sec:usecases_opt}-\ref{sec:usecases_sim} for details). 

In the remainder of this article, Sec.~\ref{sec:related_work} presents related work in terms of selected, existing benchmarking initiatives. Sec.~\ref{sec:methodology} outlines our methodology and the benchmarking metrics, while Sec.~\ref{sec:tools} describes the integration into the QUARK tooling platform. 
The selected use cases are discussed in detail across the following sections: optimization use cases in Sec.~\ref{sec:usecases_opt}, machine learning use cases in Sec.~\ref{sec:usecases_ml}, and simulation use cases in Sec.~\ref{sec:usecases_sim}. 
Finally, we conclude in Sec.~\ref{sec:conclusion}.

\subsection{Consortium}
Advancing quantum benchmarking requires a strong collaboration between leading research institutions and industrial users. The BenchQC consortium unites key players from both domains, working together within the Munich Quantum Valley (MQV) to bridge the gap between cutting-edge research and practical applications. Through this collaboration, the consortium leverages expertise in quantum computing, artificial intelligence, and optimization to address real-world challenges. The following partners contribute their specialized expertise:

The \textbf{Fraunhofer Institute for Cognitive Systems IKS} focuses on the field of safe intelligence and has vast experience in the development of reliable AI applications for markets such as autonomous driving, industry 4.0, or medical imaging. In the quantum computing domain, Fraunhofer IKS is researching QC-supported certification of classical neural networks through QC-based algorithms, and explores NISQ device applications for robust quantum machine learning.

The \textbf{Fraunhofer Institute for Integrated Circuits IIS} is a leader in international research for microelectronic and information technology solutions and services. It has extensive experience in transferring new algorithmic technologies from fundamental research to application, now expanding its focus to QC and quantum algorithms. The institute is active in various projects for mathematical optimization and quantum machine learning, including the research of the potential of QC-supported reinforcement learning in collaboration with industry partners.

\textbf{Machine Learning Reply} began exploring practical applications of quantum computing in 2018 and has implemented over 20 proof-of-concept use cases across various industries. This resulted for example in the development of an efficient classical solver for QUBO problems. ML Reply won the Airbus Quantum Computing Challenge in 2020 and reached the finals of the BMW/AWS Quantum Computing Challenge in 2021, aiming to solidify its hands-on experience through research projects.

The \textbf{OptWare GmbH} has been designing and implementing optimization and analytics in production and logistics systems for over 25 years, providing high-performance mathematical solutions for complex industry problems. The company uses AI/ML to uncover complex relationships and develops hybrid self-learning solutions that combine AI and optimization. In 2020, OptWare established an internal research group to evaluate the potential of quantum computing for its business, conducting practical studies in ongoing projects, supported by federal funding.

Vehicles are among the most complex consumer goods, with optimization potentials throughout the entire automotive value chain, particularly in production and logistics. The \textbf{BMW Group} recognized the potential of quantum computing in 2018 and established a dedicated QC team, which has since been an active member in the quantum ecosystem and was a co-founder of the Quantum Technology \& Application Consortium (QUTAC)~\cite{bayerstadler_industry_2021}. The company aims to identify when specific problem classes can benefit from quantum advantages and to translate technical advancements into business processes, addressing a key challenge for industrializing this technology.

\textbf{Quantinuum}, established through a merger of Honeywell Quantum Solutions and Cambridge Quantum Computing in 2021, builds trapped-ion quantum computers, as well as quantum software and algorithms in areas such as quantum chemistry, artificial intelligence, and quantum simulation. 
As of September 2024, Quantinuum's H2 quantum computer holds the world record in quantum volume~\cite{cross_validating_2019} at $2^{21}$.

\section{Related work}
\label{sec:related_work}
Various international quantum computing benchmarking communities have presented their strategic approaches. We here list selected examples, further overviews can be found for example in the references~\cite{riofrio_quantum_2025} or~\cite{lorenz_systematic_2025}.

The French BACQ consortium~\cite{lne_bacq_2024} brings together academic and industrial partners to benchmark quantum solutions from the realms of physics simulation, optimization, linear systems solving, and prime factorization. A special focus is put on the aggregation of application-specific metrics, such as the Q-score, into a global quality score as part of the Myriad-Q Tool. Myriad also allows for the integration of metrics from the full quantum stack.\\
Based in the United States of America, the Quantum Economic Development Consortium (QED-C) has the mission to enable industry-ready quantum computers~\cite{noauthor_quantum_nodate, noauthor_quantum_2024}. Potential analysis relies first of all on volumetric benchmarking for key functionals such as Grover's search, Hamiltonian simulation, or Shor's period finding~\cite{lubinski_application-oriented_2023}. This puts hardware-centric metrics (in particular inspired by superconducting systems), such as the quantum volume, at the center of the benchmarking process. Recently, the DARPA QB and related QBI initiative aim to verify if any quantum computing method can achieve utility-scale operation by the year 2033~\cite{darpa_qbi_2024}.\\
The Dutch TNO ecosystem, on the other hand, proposes a holistic benchmarking method called the Quantum Application Score (QuAS) to address particular user needs for a universally applicable and equitable evaluation metric~\cite{mesman_quas_2024}. 
The QuAs score represents an appropriately weighted combination of the metrics accuracy, runtime, and problem size, where the two former are taken with respect to a classical counterpart. Studied example applications from the fields of optimization and simulation include the Traveling Salesman Problem (TSP), or the Ising model.

Despite these efforts, a systematical evaluation of quantum utility in real-world scenarios is still lacking. To bridge this gap, BenchQC introduces a structured approach that emphasizes industry use cases and practical applicability. This is well aligned with the strong interest of German industry players in quantum applications, as for example represented by the QUTAC ~\cite{bayerstadler_industry_2021}.
In particular, BenchQC is uniquely characterized by the following:
\begin{itemize}
\item \textbf{Use case realism:} All studied use cases are derived from real industrial problems contributed by the corporate project partners,
\item \textbf{Hardware diversity:} The MQV develops diverse quantum hardware, for example superconducting or neutral atom systems, which will allow for representative benchmarking,
\item \textbf{Tooling:} Benchmarking builds on top of modular open-source frameworks, in particular QUARK~\cite{finzgar_quark_2022}, while maintaining compatibility with tools of the MQV ecosystem such as MQT~\cite{quetschlich_mqt_2023},
\item \textbf{Full stack approach:} Through QUARK, we establish both hardware and software interfaces for full-stack benchmarking,
\item \textbf{Metrics:} Our holistic view on quantum computing is reflected by the broad selection of metrics ranging from the hardware level up to the problem complexity on the application level,
\item 
\textbf{QML}: The utility of QML is little explored, and systematic guidance is lacking. We put a particular focus on the benchmarking of QML applications,
\item \textbf{Scalability:} Our benchmarking strategy provides various concepts to simulate problem complexity scaling.
\end{itemize}

\section{Methodology}
\label{sec:methodology}

\subsection{Problem decomposition}
\begin{figure}[h]
    \centering
    \includegraphics[width=0.45\textwidth]{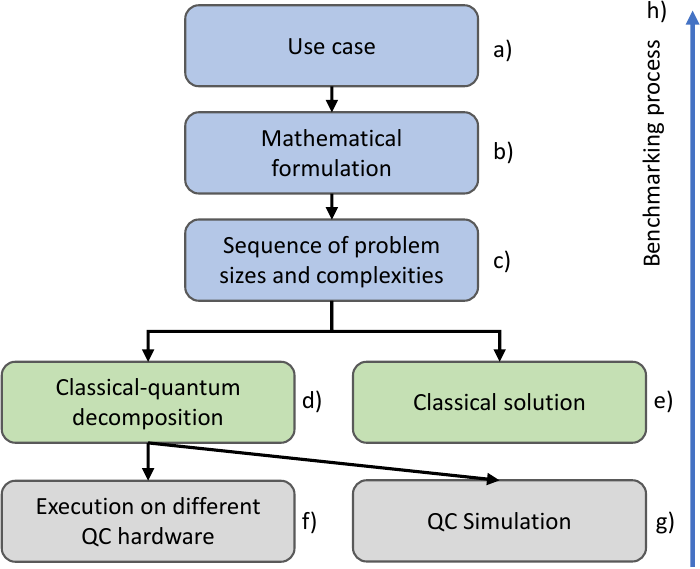}
    \caption{Decomposition of the benchmarking workflow.}
    \label{fig:workflow}
\end{figure}

To establish an application-centric benchmarking process, we follow the overall workflow illustrated in Fig.~\ref{fig:workflow}: Starting from a use case (a), we first derive a mathematical description (b). We further define a set of problem sizes and complexities (c), which together form a problem sequence to be studied. Depending on the latter, the problem is encoded as a classical-quantum hybrid algorithm (d) that can be compared to a classical reference baseline (e). The hybrid problem is eventually executed on quantum-computing hardware (f) or simulated on conventional hardware (g). Relevant metrics, as described in the following section, quantify the performance of the various parts of this decomposed workflow to govern the benchmarking process (h).

\begin{figure*}[h]
    \centering
    \includegraphics[width=0.9\textwidth]{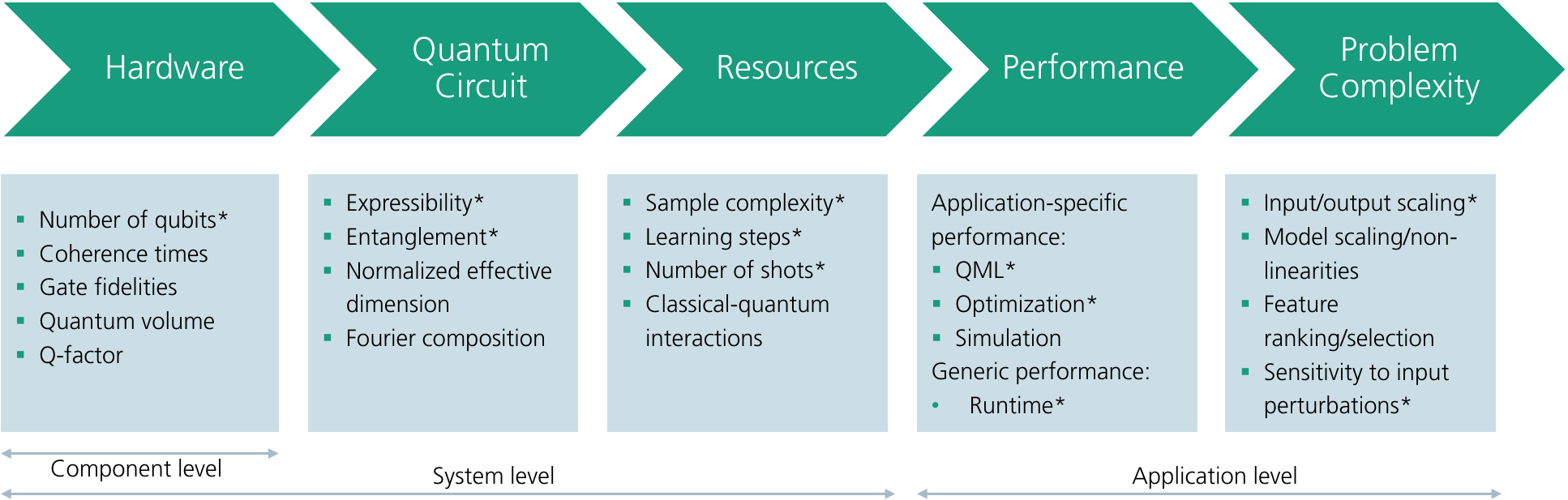}
    \caption{Metrics across the benchmarking pipeline. Asterisks (*) indicate that a form of this metric is currently implemented in QUARK.}
    \label{fig:metrics}
\end{figure*}

\subsection{Metrics}
Finding meaningful metrics to benchmark quantum computing utility is an ongoing discussion and no universally accepted framework has been established yet. Initiatives towards the standardization of quantum computing benchmarking are ongoing~\cite{lorenz_systematic_2025}. To serve as an indicator of practical quantum utility, and foster understanding of successful design patterns for quantum systems, measures need to relate the performance of an application at hand to characteristic quantum properties of the system. 
Approaches like the Q-score~\cite{martiel_benchmarking_2021} and proposed extensions~\cite{schoot_extending_2024, lne_bacq_2024} thereof evaluate the performance of a reference max-cut problem on a quantum device with respect to a random algorithm. 
Similarly, quantum volume~\cite{cross_validating_2019} or quantum \textit{Linpack}~\cite{dong_random_2021} metrics traditionally built on randomized workloads that are little representative of real use cases.
The \textit{SupermarQ} suite~\cite{tomesh_supermarq_2022} encompasses a broader set of metrics, such as entanglement or critical depth, yet falls short of enabling practical end-user applications. 
In contrast, following the spirit of exploring avenues towards relevant quantum utility, our approach in BenchQC is truly application-centric, see Fig.~\ref{fig:metrics}: Along the benchmarking pipeline, we collect a set of application-agnostic metrics, which reflect problem and solution architecture. Subsequently, we put those application-agnostic metrics in context with the actual application performance in order to identify trends for potential quantum utility. 
The metric categories are explained further in the following. 

\textbf{Hardware:}
Typical hardware metrics include coherence times (qubit lifetime $T_1$, dephasing time $T_2$, crosstalk), as well as gate fidelities~\cite{lubinski_optimization_2024, noauthor_quantum_nodate, schoot_evaluating_2023}.
The Q-factor (quality factor)~\cite{schoot_evaluating_2023} indicates the number of gate operations a qubit can perform on average before dephasing. The quantum volume (QV) is a metric popularized by the \textit{QED-C} benchmark suite~\cite{lubinski_application-oriented_2023, lubinski_optimization_2024}, which offers another hardware-based system complexity metric. It basically defines the maximum size of a random square circuit, which can be implemented with a certain tolerated error~\cite{cross_validating_2019}. The QV may, however, be less indicative of the ability of a system to achieve good solutions of practical applications~\cite{lubinski_optimization_2024}.

\textbf{Quantum circuit:}
Measuring the characteristics of quantum circuits is essential for successful parametrized quantum circuit (PQC) designs. 
First, the expressibility is the ``circuit's ability to generate (pure) states that are well representative of the Hilbert space"~\cite{sim_expressibility_2019}, which is typically a favorable characteristic for the performance of the implemented algorithm.
To measure expressibility, we use the Jensen-Shannon divergence (JSD) between the distribution of state fidelities generated by the PQC and that of an ensemble of random Haar states. Our implementation is based on the open-source library qLEET~\cite{noauthor_qleet_nodate}, yet was adapted for Qiskit v$1.1$. As the JSD is normed to one, and low values indicate higher similarity of the distributions, we use $\text{Expr}=1-\text{JSD}$ as a more intuitive expressibility metric.
On the other hand, entanglement describes the ability of a PQC to efficiently represent the solution space, i.e., capture all correlations in quantum data that might be potentially necessary to represent a quantum ground state.
For entanglement, the Meyer-Wallach (MW) multi-particle measure~\cite{sim_expressibility_2019, noauthor_qleet_nodate} is used. MW calculates the average entropy of all single-qubit reduced states, which can be interpreted as a global measure of multi-particle entanglement. A high entanglement is considered a favorable characteristic of a quantum circuit.
In particular for neural networks, the normalized effective dimension (NED) has been suggested~\cite{abbas_power_2021} as a measure for information density, and hence, for the ability of the network to learn effectively. The NED in quantum networks exceeds that of a classical counterpart and can be leveraged to derive novel generalization bounds. 
Finally, circuits with periodically repeating encoding layers can be rephrased as a Fourier series~\cite{schuld_effect_2021}. Fourier coefficients represent the circuit's expressivity and can be indicative of the model's performance~\cite{monnet_understanding_2024}.

\textbf{Resources:}
Resource metrics reflect the cost budget that is spent to derive a given solution. Relevant measures are, for example, the sampling complexity, the number of learning steps, or the number of shots of the quantum circuit. Also, the number of interactions between classical and quantum subsystems (bits or modules) is an interesting resource measure, as it represents complex operations on a real quantum device. 

\textbf{Performance:}
As a generic performance metric, the runtime or its quantum portion in a classical-quantum hybrid system is used. To assess utility, any previously discussed measures need to be linked to the performance of the application at hand. Those application-specific metrics are discussed in the individual use case chapters in Sec.~\ref{sec:usecases_opt}-\ref{sec:usecases_sim}.

\textbf{Problem complexity:}
Various strategies can be implemented to tune the complexity of a particular problem instance. Simpler problems are more likely to be compliant with NISQ limitations, yet quantum utility is expected to occur at higher complexity scales where purely classical algorithms are challenged. Practical complexity metrics such as the ones collected in Fig.~\ref{fig:metrics} can provide a means to gradually explore the interplay between both regimes.
Obvious quantitative measures to tune the problem complexity are to scale up the dimensionality of the system's input (e.g., the resolution of an image to be classified) and output (e.g., the number of possible classifications) parameters. Similarly, the model at hand can be rendered more complex by adding more connections (e.g. in a graph-like problem) or capturing possible higher-order characteristics of previously simplified model assumptions. For feature-based model designs, a varying number of input features can be selected and ranked to tune the system complexity. We further find that a simple approach to tune the problem difficulty is to pre-process the model input with artificial perturbations (e.g., by adding noise to an image).

The above categories can further be associated with component-level, system-level and application-level metrics, following the classification of~\cite{schoot_evaluating_2023, lorenz_systematic_2025}, see Fig.~\ref{fig:metrics}. Finally, any metrics aggregation approach may be employed to arrive at a global utility scale (see for instance Myriad-Q~\cite{lne_bacq_2024}) or to evaluate any selected metric with respect to a classical heuristic (e.g. QuAS~\cite{mesman_quas_2024}).

\section{The QUARK platform}
\label{sec:tools}

The Quantum Computing Application Benchmark (QUARK) framework~\cite{noauthor_quark_2025} is an open-source platform designed to evaluate the performance of quantum computing algorithms and hardware inspired by real-world scenarios. QUARK focuses on practical problems, particularly in the areas of optimization and quantum machine learning~\cite{kiwit_application-oriented_2023,kiwit_benchmarking_2024}. The platform's flexible design lets researchers add their own test scenarios, mappings, and performance metrics. Users can run benchmarks locally, on quantum simulators, or directly on quantum computers from providers like IBM~\cite{noauthor_ibm_2025}, IonQ~\cite{noauthor_ionq_2025}, and D-Wave~\cite{noauthor_d-wave_2025}.

QUARK ensures transparency and reproducibility of benchmarks by providing standardized evaluation protocols and openly accessible implementations. Its modular and flexible architecture allows seamless integration and extension of application kernels. 
Recently, the usability of QUARK has been further increased by a number of structural changes. QUARK is now available as a pip-package that manages benchmarking pipelines and provides interfaces for plugins. Several example plugins have already been provided in the framework, such as the set cover problem (SCP), a problem class that plays a crucial role in many industrial optimization tasks, e.g., the optimal configuration of LiDAR sensors within production plants. Users can add more plugins to the QUARK framework to customize their benchmarking pipelines. Each plugin provides one or more modules. The benchmarking pipeline will be constructed from modules in an implementation-agnostic way, such that a module can be connected with another module if and only if their interfaces match.

To achieve this, each module inherits the abstract \texttt{Core} module, containing a \texttt{preprocess} and \texttt{postprocess} function where the input and output types declare their interfaces. An example is shown in \cref{fig:type_hierarchy}.
Plugin authors are encouraged to choose input and output types from a set of data types provided by QUARK, to ensure compatibility with other plugins. 
If a module provides additional functionality, this can be exposed to QUARK by implementing the respective protocols (see \cref{fig:type_hierarchy}).
QUARK will then run each module sequentially while collecting performance metrics defined by the user (see~\cref{fig:quark_pipeline}).

\begin{figure}
    \centering
    \resizebox{\linewidth}{!}{
    \begin{tikzpicture}
        \umlclass[type=abstract]{Core}{}{\umlvirt{preprocess(self, data)}\\\umlvirt{postprocess(self, data)}}
        \umlclass[type=abstract, x=3.5]{Visualizable}{}{\umlvirt{visualize(self)}}
        \umlclass[type=abstract, x=-4]{Serializable}{}{\umlvirt{serialize(self)}\\\umlvirt{deserialize(cls, data)}}

        \umlclass[y=-3.5]{SCPQUBO}{lagrange\_factor}{preprocess(self, data)\\postprocess(self, data)\\serialize(self)\\deserialize(cls, data)\\visualize(self)}
        \umlVHVinherit{SCPQUBO}{Core}
        \umlHVinherit[densely dotted]{SCPQUBO}{Serializable}
        \umlHVinherit[densely dotted]{SCPQUBO}{Visualizable}
    \end{tikzpicture}
    }
    \caption{Like every QUARK module, \texttt{SCPQUBO} inherits from \texttt{Core} and implements the \texttt{preprocess} and \texttt{postprocess} functions to interface with other modules in the pipeline. By also implementing the optional \texttt{Serializable} and \texttt{Visualizable} protocols, \texttt{SCPQUBO} clearly communicates its extra capabilities to \texttt{QUARK}.}
    \label{fig:type_hierarchy}
\end{figure}
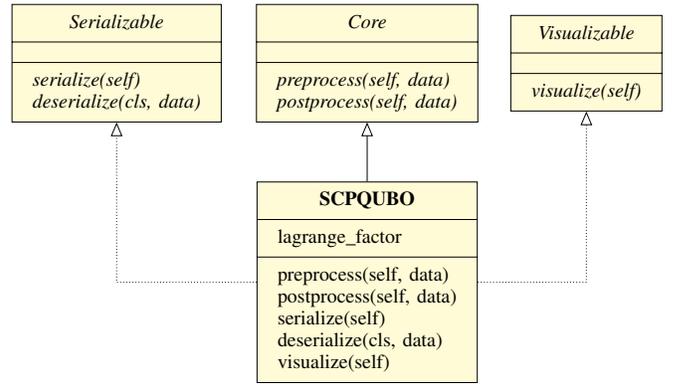

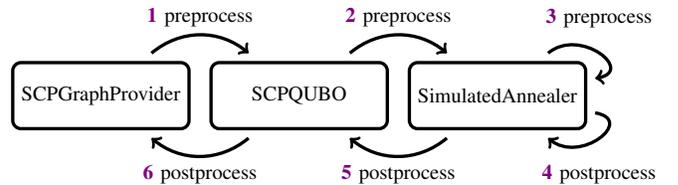
\begin{figure}
    \newlength{\radius}
    \setlength{\radius}{0.75cm}
    \tikzset{tensor/.style={fill=white, draw, circle, minimum size=2*\radius}}%
\tikzset{pipelineModule/.style={fill=white, draw, ultra thick, rectangle, rounded corners, minimum width=4*\radius, minimum height=1.5*\radius}}%
    \centering
    \resizebox{\linewidth}{!}{
\begin{tikzpicture}
    \node[pipelineModule] (SCPGraphProvider) {SCPGraphProvider};
    \node[pipelineModule] (SCPQUBO) at (4.5*\radius, 0) {SCPQUBO};
    \node[pipelineModule] (SimulatedAnnealer) at (9*\radius, 0) {SimulatedAnnealer};

    \draw[->, ultra thick] ($(SCPGraphProvider) + (40:1.5*\radius)$) to [out=40,in=140] node[above] {\textbf{\color{violet}1} preprocess} ($(SCPQUBO) + (140:1.5*\radius)$);

    \draw[->, ultra thick] ($(SCPQUBO) + (40:1.5*\radius)$) to [out=40,in=140] node[above] {\textbf{\color{violet}2} preprocess} ($(SimulatedAnnealer) + (140:1.5*\radius)$);

    \draw[->, ultra thick] ($(SimulatedAnnealer) + (40:1.5*\radius)$) to [out=40,in=10, looseness=2] node[above=3mm] {\textbf{\color{violet}3} preprocess} ($(SimulatedAnnealer) + (10:2.25*\radius)$);

    \draw[->, ultra thick] ($(SimulatedAnnealer) + (-10:2.25*\radius)$) to [out=-10,in=-40, looseness=2] node[below=3mm] {\textbf{\color{violet}4} postprocess} ($(SimulatedAnnealer) + (-40:1.5*\radius)$);

    \draw[->, ultra thick] ($(SimulatedAnnealer) + (220:1.5*\radius)$) to [out=220,in=320] node[below] {\textbf{\color{violet}5} postprocess} ($(SCPQUBO) + (320:1.5*\radius)$);

    \draw[->, ultra thick] ($(SCPQUBO) + (220:1.5*\radius)$) to [out=220,in=320] node[below] {\textbf{\color{violet}6} postprocess} ($(SCPGraphProvider) + (320:1.5*\radius)$);
\end{tikzpicture}
}
\caption{The QUARK benchmarking pipeline is made up of modules that can pre- and postprocess data. In this case, the set cover problem is solved in two phases: By calling the preprocess functions of each module in order, a graph is created \textbf{(1)}, mapped to a QUBO formulation \textbf{(2)}, and solved on a simulated annealer \textbf{(3)}. Afterwards, calling the postprocess functions of each module in reverse order, the lowest energy sample is obtained \textbf{(4)}, re-interpreted as a list of nodes \textbf{(5)}, and compared with the graph \textbf{(6)}.}
\label{fig:quark_pipeline}
\end{figure}


Further details and instructions on the use of QUARK can be found in the developer guide~\cite{noauthor_quark_2025}.
In BenchQC, we leverage QUARK as a platform to study realistic use cases.
The following sections give a review of selected problems - along with their strengths and challenges - from each of the studied categories of optimization, machine learning, and simulation. A full list is given in Tab.~\ref{tab:use_cases}.

\begin{table*}[h]
    \caption{Table of use cases in BenchQC.}
    \label{tab:use_cases}
    \centering
    \begin{tabular}{|c|c|c|}
        \hline
        \textbf{Category} & \textbf{Model} & \textbf{Use case} \\
        \hline
        \multirow{5}{*}{QML} & Classification & Defect detection \\ \cline{2-3}
        & \multirow{2}{*}{Generative models} & Generative design for constructional elements \\ \cline{3-3}
        & & Data augmentation for quality inspection\\ \cline{2-3}
        & \multirow{1}{*}{QRL} & 6G Beam management\\ 
        \hline
        \multirow{6}{*}{Optimization} & \multirow{6}{*}{VQE, QAOA, Quantum Annealing} & Multipath-connectivity across aerial vehicles \\  \cline{3-3}
        & & Sensor placement in water networks \\ \cline{3-3}
        & & Software verification strategies\\  \cline{3-3}
        & & LiDAR sensor configuration \\ \cline{3-3}
        & & Assembly line scheduling and packet sequencing\\ \cline{3-3}
        & & Financial asset optimization\\
        \hline
        \multirow{3}{*}{Simulation} & \multirow{3}{*}{Hamiltonian simulation} & Dynamics of electrons in materials \\  \cline{3-3}
        & & Low-temperature state preparation of materials\\ \cline{3-3}
        & & Nuclear magnetic resonance simulation\\  \cline{3-3}
        \hline
    \end{tabular}
\end{table*}

\section{Use cases: Optimization}
\label{sec:usecases_opt}

\subsection{Basics of optimization}
\label{sec:opt_basics}
In general, mathematical optimization refers to the process of finding the best solution within a defined set of feasible solutions. An optimization problem can be expressed as
\[
\text{Minimize } f(x) \quad \text{subject to } x \in S,
\]
where \( f(x) \) is a real-valued function representing the objective, and \( S \) denotes the set of feasible solutions.
Such problems arise in various fields, including operations research, machine learning, industry and finance~\cite{williams_model_2013, pochet_production_2006, cornuejols_optimization_2018}.

The solution methods for these problems are diverse.
For Mixed-Integer-Linear-Programs (MIP), one prominent approach is \textit{Branch and Bound (B\&B)}, which explores subsets of the solution space, using bounds to exclude suboptimal branches effectively~\cite{land_automatic_1960}.
Another common method to solve MIPs is the \textit{Cutting Plane Method}, where the feasible region is iteratively refined by introducing linear constraints, or ``cuts," to exclude infeasible regions while preserving all valid solutions\cite{marchand_cutting_2002}.
State-of-the-art solvers implement so-called \emph{branch-and-cut} algorithms, which effectively combine B\&B and cutting plane methods.
Additionally, \textit{heuristic methods}, such as Genetic Algorithms and Simulated Annealing, can provide approximate solutions with reduced computational complexity~\cite{conforti_integer_2014}. 

Both open-source solvers like CBC~\cite{noauthor_coin-orcbc_2025}, GLPK~\cite{noauthor_glpk_nodate}, HIGHs~\cite{noauthor_highs_nodate} and SCIP~\cite{bestuzheva_enabling_2023}, as well as commercial solvers such as CPLEX~\cite{noauthor_solver_2024} and Gurobi~\cite{noauthor_leader_nodate}, are widely used to tackle MIPs.
For other types of optimization problems, e.g., with non-linearities, there exist other methods like Satisfiability Modulo Theories (SMT) solvers~\cite{de_moura_satisfiability_2011} or Semidefinite Programming (SDP) solvers~\cite{wolkowicz_handbook_2000} that are better suited.
Certain classes of optimization problems are particularly challenging to solve. Quadratic programming problems, characterized by quadratic objectives or constraints, often pose significant computational difficulties~\cite{furini_qplib_2019}. Similarly, non-linear programming problems and non-convex problems, which involve multiple local optima, are notoriously hard to solve, as finding the global optimum requires advanced and resource-intensive methods\cite{aragon_nonlinear_2019}. In cases where classical solvers reach their limits, quantum methods emerge as a potential means to address these computationally demanding optimization problems.

\subsection{Benchmarking Quantum Computing Optimization in the NISQ Era}
In the NISQ era, quantum computing has emerged as a tool for addressing combinatorial optimization problems that can possibly have a high impact in the near future. Quantum algorithms, such as the Quantum Approximate Optimization Algorithm (QAOA)~\cite{farhi_quantum_2014} and Quantum Annealing~\cite{hauke_perspectives_2020}, leverage binary formulations of these problems to utilize current quantum hardware.
Such algorithms represent promising avenues towards solving large-scale optimization problems in fewer computation steps than their classical counterparts~\cite{wybo_missing_2024, bauza_scaling_2024, shaydulin_evidence_2024}.
However, the current limitations of quantum hardware restrict their application to relatively small-scale problems, underscoring the need for further advancements in hardware capabilities~\cite{guerreschi_qaoa_2019, junger_quantum_2021}.
As hardware improves, it is essential to simultaneously advance software development to ensure that quantum algorithms are ready for deployment on next-generation quantum devices. 

\subsection{Metrics for Benchmarking Quantum Optimization}
Using benchmarking, we enable informed decisions about the potential of quantum computing for a given algorithmic problem~\cite{acuaviva_benchmarking_2024}.
One fundamental metric for multi-stage quantum algorithms like QAOA is the \textit{runtime}, typically given as a separate assessment of the parameter optimization time and the execution time of the quantum circuit~\cite{lubinski_application-oriented_2023}.
The application-specific performance is measured by a \textit{solution quality} metric. 
Solution quality of the quantum method can be assessed by calculating the absolute and relative optimality gap (\( \Delta_{abs} \) and \( \Delta_{rel} \)) and the approximation ratio \( \Theta \), if the optimal solution is known
\begin{equation}
\Delta_{abs} = f_\text{qc} - f_\text{opt}, \quad
\Delta_{rel} = \frac{f_\text{qc} - f_\text{opt}}{f_{opt}}, \quad
\Theta = \frac{f_\text{qc}}{f_\text{opt}},
\end{equation}
where \( f_\text{qc} \) is the objective value of the quantum solution, and \( f_\text{opt} \) is the value of the optimal solution, assuming \( f_\text{opt}, f_\text{qc} \geq 0 \). When the optimal solution is unknown, solution quality can be quantified through alternative methods, such as comparing quantum solutions to those found by classical algorithms within the same runtime or benchmarking them against random sampling techniques.

The \textit{Q-Score}~\cite{martiel_benchmarking_2021} provides a combined evaluation of computation time and solution quality. For MaxCut problems, it defines the problem size \( N \) at which a quantum algorithm outperforms a random algorithm within a $60$-second limit. The fundamental metric for the Q-Score is calculated as follows:
\begin{equation}
\beta(N) = \frac{C(N) - C_\text{rand}(N)}{C_\text{opt}(N) - C_\text{rand}(N)},
\end{equation}
where \( C(N) \) is the average best cut found by the quantum algorithm, \( C_\text{opt}(N) \) is the average optimal cut size, and \( C_\text{rand}(N) \) is the average cut size of random solutions. As proposed by Atos~\cite{martiel_benchmarking_2021}, an arbitrary threshold value of \( \beta(N) \geq 0.2 \) is used to determine if for problem size N the quantum algorithm still outperforms a random algorithm. 
The Q-Score is adaptable to other problems, as demonstrated for MaxClique~\cite{schoot_extending_2024}, see also our use case in the following section.

\subsection{Use Case: Portfolio Optimization}

Originally formulated by Harry Markowitz~\cite{markowitz_portfolio_1952}, the portfolio optimization problem asks for a selection of $n$ possible assets from an asset portfolio, which maximizes the return $\mu$ while minimizing the return variance $\sigma^2$, also called \emph{volatility}.
Mathematically, these objectives can be expressed as
\begin{equation}
\mu = \sum_{i=1}^n w_i r_i, \qquad \sigma^2 = \sum_{i=1}^n \sum_{j=1}^n w_i w_j \sigma_{ij},
\end{equation}
where the variables \( w_i \) represent the asset weights that must sum up to one.
The parameters \( r_i \) are the expected returns and \( \sigma_{ij} \) are the covariances between assets.
Portfolio optimization thus constitutes a multi-objective optimization problem.
Due to the quadratic nature of the return variance, classical solvers struggle to scale efficiently with an increasing number of assets, making quantum computing a natural consideration for solving this problem.

Various formulations of the portfolio optimization problem are studied in the literature, including maximizing return with constrained volatility (``Maxret") as in~\cite{sakuler_real_2023},~\cite{palmer_quantum_2021} , minimizing volatility with constrained return (``Minvola") as in~\cite{venturelli_reverse_2019},~\cite{cesarone_efficient_nodate}, and maximizing a trade-off between return and volatility \( \mu - \lambda \sigma^2 \) (``Multiobj"), where \( \lambda \) is a parameter balancing the two objectives~\cite{brandhofer_benchmarking_2022},~\cite{acharya_decomposition_2024},~\cite{rubio-garcia_portfolio_2022}. For a comprehensive overview on portfolio optimization we refer the interested reader to~\cite{loke_portfolio_2023}.
Our analysis of the three formulations above yields that the Minvola problem formulation is the most difficult to solve with classical optimization methods, as can be seen from the longer classical solution times of the Minvola formulation compared to the Maxret and Multiobj formulation in Fig.~\ref{fig:solution_times}. This makes it the natural choice to investigate possible benefits of quantum-assisted solutions.
\begin{figure}[h]
    \centering
    \makebox[\textwidth][l]{ 
        \includegraphics[width=0.98\linewidth]{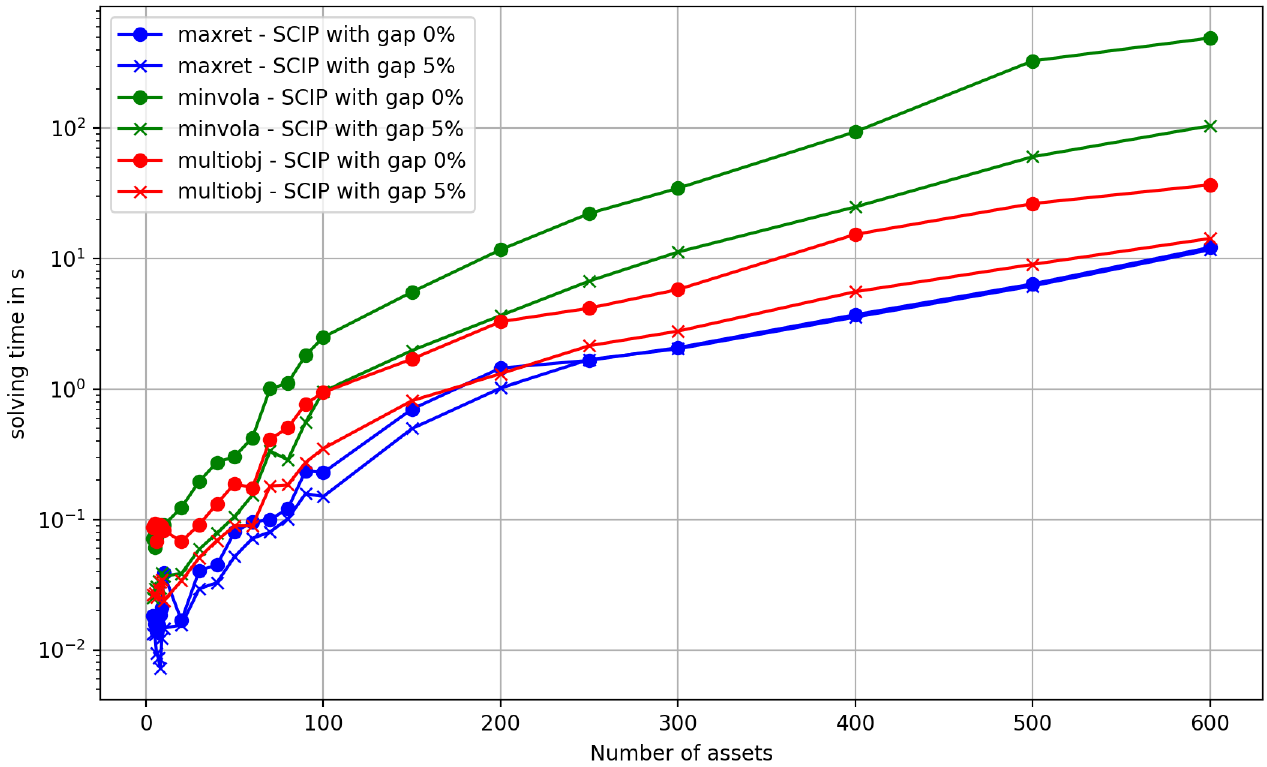}
    }
    \caption{\label{fig:solution_times}Runtime comparison for solving the Minvola, Maxret and Multiobj problem formulations of the portfolio optimization problem with SCIP~\cite{bestuzheva_enabling_2023} with relative optimality gaps of $0\%$ and $5\%$.
    }
\end{figure}

The research approach in this project involved the evaluation of classical solvers and heuristics on this formulation to assess their limitations. 
For future work, a large-scale benchmarking study will be conducted to determine the quantum potential for solving the Minvola formulation. 
For problem instances involving a given number of assets, multiple problems will be randomly generated from a set of approximately 2000 assets from real-world financial data. 
Each problem will be solved using classical solvers, Quantum Annealing, QAOA, and classical heuristics. 
For the quantum methods, only problems up to a certain asset size can be solved, thus the boundaries of Quantum Annealing and QAOA will be tested during this study.
Metrics like optimality gap, approximation ratio, an adaptation of the Q-Score and others will be used to evaluate performance, with results averaged across all instances of a certain problem size to derive robust conclusions.

\subsection{Use Case: Assembly Line Balancing}

Modern manufacturing often means assembly line mass production in modern factories. Products are completed stepwise in a sequence of multiple workstations, numbered~$s \in \mathbb{S} \subset \mathbb{N}$ in line order. Synchronized by a common cycle time~$c \in \mathbb{R}^+$, each workstation performs a subset of specific assembly tasks~$t \in \mathbb{T}$.

The \textit{Simple Assembly Line Balancing Problem}~\cite{scholl_balancing_nodate} describes the planning process of distributing each task to exactly one workstation (see Eq.~\eqref{eq:salbp_2}), minimizing the number of stations (Eq.~\eqref{eq:salbp_1}), while respecting task precedence relations~$(t, t') \in \mathbb{E} \subset \mathbb N \times \mathbb N $ induced by manufacturing order, which demands that $t$ must preceed $t'$ (Eq.~\eqref{eq:salbp_4}). Each task is assigned an expected mean processing time $v_t \in \mathbb{R}$, multiplying time for task completion by the mean frequency of task execution. The summed processing time for each station must not exceed the cycle time (Eq.~\eqref{eq:salbp_3}).

Using the binary decision variables $x_{ts}$ (``true" if task $t$ is on station $s$), and $y_{s}$ (``true" if station $s$ is needed), the problem takes the form:
\begin{align}
   \text{min} & \sum_{s} s \cdot y_s & \label{eq:salbp_1}\\
   \text{s.t.} & \sum_{s \in S} x_{ts} = 1 & \forall t \in T \label{eq:salbp_2} \\
        & \sum_{t \in T} v_t \cdot x_{ts} \leq c \cdot y_s &\forall s \in S \label{eq:salbp_3} \\
        & \sum_{s \in S} s \cdot x_{ts} \leq \sum_{s \in S} s \cdot x_{t's} &\forall (t, t') \in \mathbb{E} \label{eq:salbp_4} \\
        & x_{ts}, y_s \in \{0,1\} & \forall t \in T, s \in S \label{eq:salbp_6} 
\end{align}
This MIP formulation is suitable for classical optimization as described in Sec.~\ref{sec:opt_basics}. In order to solve this problem using Quantum Annealing or QAOA, the MIP formulation has to be mapped to a QUBO formulation by transforming the constraints Eqs.~\eqref{eq:salbp_2}-\eqref{eq:salbp_6} into penalties that extend the objective function (Eq.~\eqref{eq:salbp_1}). This transformation can be done using slack variables, e.g. implemented in Qiskit~\cite{javadi-abhari_quantum_2024}, or using unbalanced penalization~\cite{montanez-barrera_unbalanced_2024}. 
In practice, problem instances reach complexities where classical solvers struggle to produce high-quality solutions at a reasonable runtime, and hence a quantum advantage could become relevant in the future.
Currently, studies will be conducted using instances that can be easily scaled down in complexity to remain tractable on existing quantum hardware.

\section{Use cases: Machine learning}
\label{sec:usecases_ml}

\subsection{Use case: QRL for beam management}
In this section, we study a use case derived from a task known as handover management in 6G wireless communication~\cite{corici_fraunhofer_2021} (cf. Tab.~\ref{tab:use_cases}). 
Next-generation communication networks will feature antennas
that can generate directional beams to serve users, e.g. mobile phones. The task of beam management is to identify the antenna and beam direction that maximize beam quality at the position of user. The antenna indices together with the discretized beam directions are pre-coded in a so-called codebook. Without 
knowledge of the exact trajectory of the mobile phone, a reinforcement learning (RL) agent is trained to optimally select the antenna index and codebook element based on previous choices. For a detailed description of the use case, we refer to~\cite{meyer_benchmarking_2025}. A general toy-model simulator~\cite{meyer_6g_2025} is available open-source. Unlike typical community benchmarks~\cite{brockman_openai_2016,tassa_deepmind_2018}, this simulator allows for flexible scaling of the environment and is particularly suited for quantum computing due to its small input and output dimensions~\cite{hoefler_disentangling_2023}.

Reinforcement learning~\cite{sutton_reinforcement_2018} is a subfield of machine learning, where an agent (i.e. the RL algorithm) learns a decision policy by trial and error through interactions with an environment (i.e. the problem to solve). This algorithmic paradigm is increasingly applied in industry~\cite{arulkumaran_deep_2017, francois-lavet_introduction_2018} and has recently seen a surge of interest for fine-tuning large language models\cite{ziegler_fine-tuning_2020} and training early reasoning models~\cite{deepseek-ai_deepseek-r1_2025}. Several quantum versions of RL, known as quantum reinforcement learning (QRL)~\cite{meyer_survey_2022}, have been proposed. Some of these versions provably outperform classical RL algorithms~\cite{jerbi_parametrized_2021, liu_rigorous_2021} in artificial tasks without practical relevance or promise quantum advantage in comparison to specific classical RL
\cite{wang_quantum_2021, ganguly_quantum_2023, zhong_provably_2023,cherrat_quantum_2023,dunjko_quantum-enhanced_2016} algorithms with limited significance in industrial settings.
Since contemporary RL algorithms typically utilize deep neural networks as function approximators, we focus on quantum versions of these models. Here, the conventional deep neural network is substituted~\cite{chen_variational_2020, jerbi_parametrized_2021, skolik_quantum_2022} with a parametrized 
quantum circuit~\cite{bharti_noisy_2022}. In analogy to their classical counterparts, theoretical understanding of the performance of these quantum models remains limited. As a result, the search for evidence of quantum advantage for this type of QRL is presently largely an empirical endeavor. 

\subsection{Benchmarking QRL}
In industrial settings, data acquisition can be costly, in particular when dealing with edge cases. Consequently, the data efficiency of the RL algorithm is a critical concern. We therefore propose to benchmark QRL with respect to a formalized notion of this intuition, referred to as \textit{sample complexity}~\cite{kearns_finite-sample_1999, kakade_sample_2003} and defined as the number of interactions with the environment in the training process, required to reach a certain performance threshold with high probability. This metric is listed in Fig.~\ref{fig:metrics} in the \textit{Resources} category. Performance of different QRL models on further metrics have recently been investigated in ~\cite{kruse_benchmarking_2025}.
Previous work~\cite{chen_variational_2020, dragan_quantum_2024, reers_towards_2023, hohenfeld_quantum_2024} evaluating sample complexity has made claims asserting the superiority of QRL over classical RL on this metric. However, such claims must be taken with great care, as the training process is influenced by many sources of randomness, such as stochasticity in the environment, random initialization of network parameters, and shot noise~\cite{henderson_deep_2018, jordan_position_2024, bowles_better_2024,franz_uncovering_2023}. In our recent work~\cite{meyer_benchmarking_2025}, we therefore strongly advocate robust statistical evaluation procedures to avoid false or statistically insignificant claims.
To this end, we formalized the sample complexity metric in terms of a statistical estimator backed by significance testing. Utilizing fast simulation libraries~\cite{bergholm_pennylane_2022, meyer_qiskit-torch-module_2024}, we performed 100 independent training runs per environment instance to achieve statistical significance, by far the largest number of runs performed in this context to date.
\begin{figure}[t]
    \centering
    \includegraphics{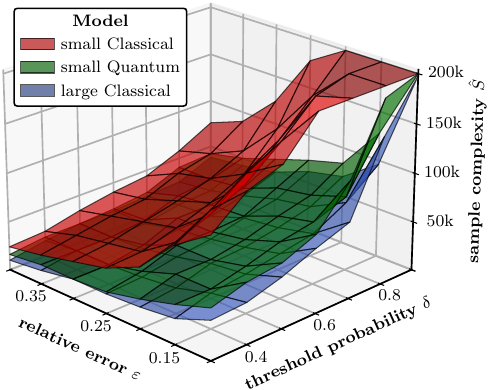}
    \caption{\label{fig:sample_complexity}Comparison of empirical sample complexities of double deep Q-learning and a quantum version of the algorithm (lower is better).  In order of decreasing sample complexity: a \emph{small classical} neural network with $2$ hidden layers of width $16$, i.e., $387$ parameters; a \emph{small quantum} circuit with $4$ layers on $14$ qubits, i.e., $437$ variational parameters, integrated between fully connected classical layers with additional $101$ parameters; a \emph{large classical} neural network with $2$ hidden layers of width $64$, i.e., $4611$ parameters. Figure taken from Ref.~\cite{meyer_benchmarking_2025}.
    }
\end{figure}
Fig.~\ref{fig:sample_complexity} compares the sample complexity between classical double Q-learning~\cite{van_hasselt_deep_2016} and a quantum version of the algorithm that employs parameterized quantum circuits as function approximators for a specific environment instance. The figure shows the sample complexity as a function of $\epsilon$ and $\delta$. Here, $1-\epsilon$ is the ratio of the chosen performance threshold to the ground truth optimal solution, and $\delta$ is the confidence probability. 
We first choose the best performing classical and quantum model type based on an extensive architecture and hyperparameter search and then define a sequence of classical and quantum models based on the number of trainable parameters. 
This one-dimensional projection in model space serves as a viable proxy for model capacity. The figure illustrates the typical behavior observed across many environments and algorithmic realizations. Notably, the quantum model's performance is almost on par with much larger classical models. When further increasing the quantum model width (the number of qubits) as well as the parameter count of the classical model, the performance of the quantum model continues to improve, while the classical model tends to reach a saturation point. This may indicate a trend that warrants further investigation which will be the focus of future work.

\subsection{Use case: QCNN for defect detection}
Also for image classification tasks, quantum convolutional neural networks (QCNNs) represent a promising application of quantum computing~\cite{abbas_power_2021, zhao_review_2021, monnet_understanding_2024}. We investigate how hybrid quantum-classical approaches can enhance a defect detection system, in particular, a classification model on a specialized dataset of images of road surface segments with and without surface cracks~\cite{bmw_road_2021} (see Fig.~\ref{fig:qcnn_figure} for an example).

Our QCNN implementation follows an architecture that seeks to combine the strengths of classical deep learning and quantum computation. The system consists of the following layers: 
\begin{enumerate}
    \item \textbf{Feature extraction:} A pre-trained classical ResNet-18 CNN~\cite{he_deep_2015} processes the input images to extract features,
    \item \textbf{Dimensionality reduction:} A classical fully connected layer reduces features to match the available quantum resources,
    \item \textbf{Quantum circuit:} A parameterized quantum circuit that processes these compressed features,
    \item \textbf{Classification layer:} A classical output layer for final classification.
\end{enumerate}
The quantum portion of the network employs a structured circuit making use of a strongly entangling layer~\cite{schuld_circuit-centric_2020, monnet_understanding_2024}, following a set of Hadamard gates for initial superposition. Such an encoding strategy allows the quantum circuit to explore complex feature interactions, that may be challenging to represent classically, even with a small number of qubits on limited NISQ hardware. 
At the same time, using classical pre-processing, the hybrid network can handle input data of realistic sizes. 

%

\begin{table}[ht]
    \caption{Accuracy of classical and classical-quantum hybrid models on the road crack dataset with a varying number of input features. $1024$ training images were used per label.}
    \centering
    \begin{tabular}{| p{2cm} | p{2cm} | p{2cm} |}
        \hline
         \# Features after dimensionality reduction (qubits) & Classification accuracy of the hybrid model & Classification accuracy of the classical model \\
         \hline
         2 & - & 0.999542 \\
         4 & - & 0.999635 \\
         6 & 0.996825 & 0.999645 \\
         8 & 0.999359 & 0.999037 \\
         \hline
    \end{tabular}
    \label{tab:hybridaccuracy}
\end{table}

\subsection{QCNN Complexity Analysis and QUARK Implementation}
We have tested the scalability of the hybrid architecture by varying several key parameters, namely the number of qubits ($2$, $4$, $6$, and $8$ qubits), the dimension of the input feature space, and circuit depth.
Our experiments show that hybrid models using $6-8$ qubits achieve comparable performance to their classical counterparts, with classification accuracies $>0.996$ each, see Tbl.~\ref{tab:hybridaccuracy}. 

To comprehensively evaluate quantum machine learning models, we employ various metrics selected from our analysis in Sec.~\ref{sec:methodology}.
Classical performance metrics indicate how a quantum machine learning model holds up against the classical counterpart. Examples include classification accuracy on test and training sets, area under curve (AUC) of receiver operating characteristics (ROC) and comparative training efficiency measuring convergence time. 
Quantum circuit metrics, on the other hand, are employed to assess the quantum properties of the hybrid system. In the QCNN use case, we make use of the entanglement and expressibility. The ratio between the quantum part and the overall execution time shows the compute efficiency. Finally, quantum hardware characteristics are given by the number of qubits and circuit depth.
%

Our integration with the QUARK framework enables systematic collection and analysis of these metrics across different model configurations. 
This allows for standardized evaluation and comparison with other quantum applications. For easy interpretation and subsequent aggregation schemes, we provide a comprehensive view of selected performance metrics, see also Fig.~\ref{fig:qcnn_figure}.


\subsection{Defect detection: Results}
\begin{figure}[t]
    \centering
    \includegraphics[width=0.4\textwidth]{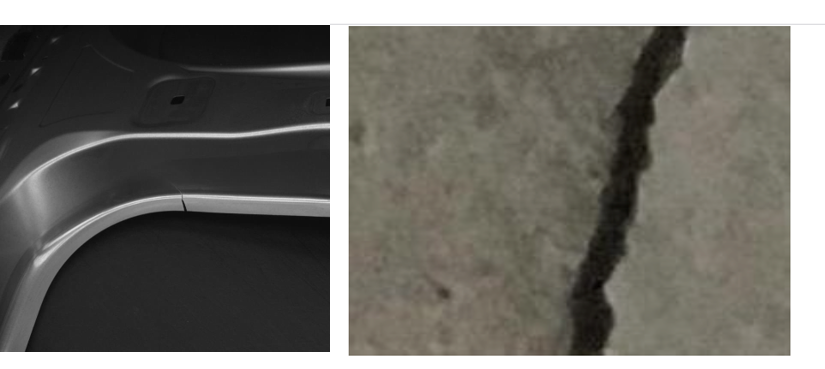}\\
    \includegraphics[width=0.4\textwidth]{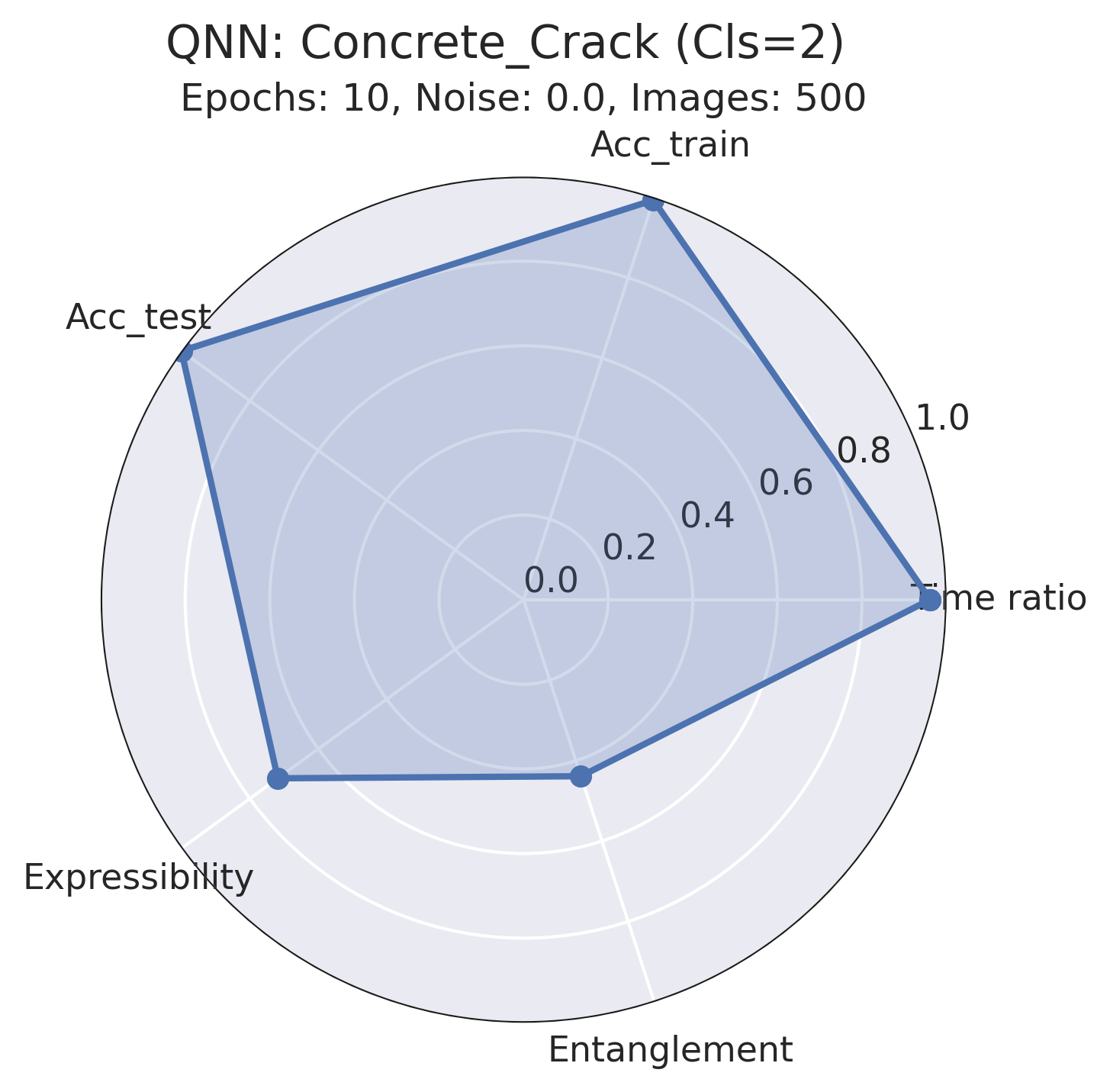}
\caption{
Top: Example images of the road crack dataset~\cite{bmw_road_2021}. Bottom: Selected key metrics of the QCNN trained on that dataset for defect detection. The time ratio measures the portion of execution time on the quantum circuit compared to the overall execution time. The neural network performance is measured as train and test accuracies.}
\label{fig:qcnn_figure}
\end{figure}

In the given setup, a comparison of the hybrid and classical models reveals several key insights (cf with Fig.~\ref{fig:qcnn_figure} and Tab.~\ref{tab:hybridaccuracy}): 
\begin{itemize}
    \item For the concrete crack classification task, both classical and quantum approaches achieve high accuracy ($>99.5\%$) on the train and test datasets. Similarly, AUC values for both approaches reach $>0.996\%$ for $6$ features,
    \item The quantum model showed slightly more variable training convergence, with initial higher loss values but eventual convergence comparable to the classical model,
    \item Entanglement measurements showed a relatively low value ($\approx 0.4$), indicating limited quantum entanglement in our circuit design,
    \item Expressibility metrics revealed a moderately high ($\approx0.7$), suggesting that our quantum circuit might not be utilizing its full expressive power,
    \item Time ratios close to one indicate that most of the computational time was spent on the execution of the quantum layers.
\end{itemize}
The results indicate that - while quantum approaches do not currently offer a clear advantage in terms of raw classification performance for this task - they demonstrate competitive capabilities that warrant further exploration as quantum hardware improves. 
The entanglement and expressibility results are particularly informative for quantum circuit design. Despite achieving high classification accuracy, the relatively low entanglement value suggests potential room for improvement in leveraging quantum correlations. Similarly, the expressibility score indicates that our current circuit architecture may not fully explore the available Hilbert space. These quantum-specific metrics provide valuable guidance for circuit optimization beyond traditional accuracy measures. 

\section{Use cases: Simulation}
\label{sec:usecases_sim}

\subsection{Use Case: Dynamic of electrons in materials} 
Simulation of many-body quantum systems is one of the simplest tasks that quantum computers can perform exponentially faster than any known classical algorithms~\cite{lloyd_universal_1996,kim_evidence_2023,haghshenas_digital_2025}. It is used in material science~\cite{ma_quantum_2020}, quantum chemistry~\cite{cao_quantum_2019} and condensed matter physics~\cite{kim_evidence_2023}, which in turn find applications in semiconductor technology, energy storage, pharmaceuticals and aerospace. In quantum computing, simulation appears as a sub-routine in many applications, such as in quantum phase estimation to determine ground state energy of molecules and reaction rates. Its direct industrial applications range from drug discovery to material discovery, enabling efficient simulations of crucial experiments such as nuclear magnetic resonance or neutron-scattering experiments.

When materials are probed with techniques such as neutron-scattering experiments, angle-resolved photoemission spectroscopy or nuclear magnetic resonance, the signal measured by the device can be expressed in terms of dynamical correlations of observables in the system. To elucidate the structure of the materials, one has to simulate the dynamics of an underlying Hamiltonian and compare the simulation to the experiments. Some quantum computers themselves are built out of fermionic ions or neutral atoms that undergo complex dynamics. As a representative use case to include in this benchmark suite, we consider the simulation of the dynamics of fermions. This use case encompasses an array of Hamiltonian features and demonstrates representative computational complexity. A concrete use case with wide industrial and academic interest is the Hubbard model~\cite{agrawal_quantifying_2024}, which can be used to describe high-temperature superconductivity in cuprates~\cite{arovas_hubbard_2022}. It is mathematically formulated as
\begin{equation}
    H=-t\sum_{\langle i,j\rangle,\sigma} (c_{i,\sigma}^\dagger c_{j,\sigma}+c_{j,\sigma}^\dagger c_{i,\sigma})+V\sum_{i}\left(n_{i,\uparrow}n_{i,\downarrow}-\frac{1}{4}\right)\,,
\end{equation}
where $c^\dagger,c$ denotes fermionic creation and annihilation operators, $\sigma\in \uparrow,\downarrow$ their spin, $n_{i,\sigma}$ the density of fermions, $t,V$ parameters, and $\langle i,j\rangle$ that sites $i,j$ are neighbours on a lattice. The classical computational cost to simulate the time evolution of this system is exponential in either system size for state-vector methods, or in simulation time for tensor networks~\cite{orus_tensor_2019}, Pauli string expansion~\cite{schuster_polynomial-time_2024} or neural network methods~\cite{carleo_solving_2017}. The case $V=0$ is however useful for benchmarking purposes, as it can be classically simulated in polynomial time, while still involving very similar circuits to the case $V\neq 0$, non-Clifford gates and large entanglement. By time evolving an initial state and measuring an observable, one can evaluate the accuracy of quantum computing hardware in a scalable way. Varying the simulation time varies multiple features of the hardware panel of Fig.~\ref{fig:metrics}, such as the number of qubits and gate fidelity. We give in Ref.~\cite{granet_appqsim_2025} a detailed description of the benchmarking protocol. The performance of the tested hardware can be evaluated with metrics of Fig.~\ref{fig:metrics} with e.g. runtime or number of shots to reach a given precision. We also define a specific performance metric for this use case called the \textit{distinguishability cost}, that is, given the output of a tested hardware, the minimal number of gates to be run on an ideal perfect quantum computer (possibly across different shots) to be able to detect an inaccuracy in the output~\cite{granet_appqsim_2025}. This number increases with better hardware quality, in the sense that a noisy quantum computer with significant bias will require fewer shots to be distinguished from a noiseless quantum computer. It balances device runtime and gate precision, because the score of a very accurate but slow hardware will be bounded by the number of shots it can run. It is applicable both to actual quantum hardware to measure hardware noise, and to classical simulation methods to evaluate the relevance of quantum computers for this task. An illustration of this use case is shown in the left panel of Fig~\ref{fig:simulation}.

\begin{figure}[h]
    \centering
    \includegraphics[width=0.49\linewidth]{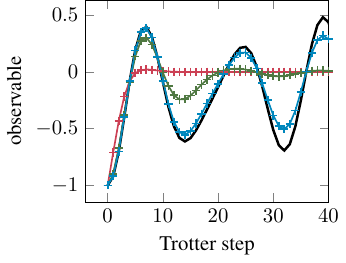}
\includegraphics[width=0.49\linewidth]{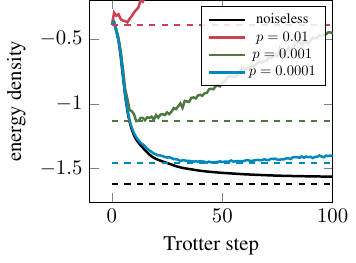}
    \caption{\label{fig:simulation}Left: fermion imbalance as a function of number of Trotter steps for the dynamic use case, in a free fermion system with compact encoding of size $4\times 4$, for different error rates $p$ per two-qubit gates, with the labels displayed in the right panel. Right: energy density as a function of number of Trotter steps for the static use case, for the Heisenberg model in a Kagome lattice with $18$ sites, for different error rates. The dashed lines indicate the minimum attained for each.
    }
\end{figure}

\subsection{Use Case: Low-temperature state preparation of materials}
The other application of quantum computers in many-body simulation that we consider is the computation of \emph{static} properties, such as ground-state or finite-temperature observables~\cite{nigmatullin_experimental_2024}. This finds direct applications in condensed-matter physics and material discovery, for example in the study of superconductivity in cuprates, where the first step of any experiment is the preparation of a low-temperature state. A systematic approach to prepare low-energy states is the adiabatic algorithm~\cite{albash_adiabatic_2018}, also called quantum annealing in the classical optimization context. Generally, equilibrium properties are almost as difficult to compute classically as dynamical properties. The preparation of a low-temperature state and the measurement of its energy can be used as a scalable benchmark, as different hardware or classical methods can be compared beyond the exactly simulatable regime. Moreover, this task is known to display a low sensitivity to errors comparable to measuring local observables instead of global state fidelity~\cite{granet_dilution_2025,chertkov_robustness_2024}. We will take as a use case the Heisenberg model on a Kagome lattice, which describes the material ${\rm YCu}_3 {\rm [OH(D)]}_{6.5}{\rm Br}_{2.5}$~\cite{liu_gapless_2022}. The Hamiltonian is
\begin{equation}
    H=-\sum_{\langle i,j\rangle}X_iX_j+Y_iY_j+Z_iZ_j,
\end{equation}
where $\langle i,j\rangle$ means that $i,j$ are neighbours on a Kagome lattice. This system is known to be classically challenging and has unsolved open physical questions. The problem can be scaled to arbitrary system sizes. Starting from a given initial state, we define a score as being the lowest energy attained when implementing an adiabatic evolution with a fixed path and schedule, that we detail in Ref.~\cite{granet_appqsim_2025}. The effect of hardware noise in these systems is known to be low, and the precision obtained on the energy is a good proxy for the precision obtained on typical local observables~\cite{granet_dilution_2025,chertkov_robustness_2024}. To accommodate both near-term and longer-term devices, we define two different initial states to be run on NISQ and post-NISQ devices. We show numerical simulations of this use case in the right panel of Fig.~\ref{fig:simulation} for the NISQ initial state. Even beyond the classical simulability of the adiabatic evolution, two different hardware devices can be evaluated by comparing the energy of the states they can prepare.

\section{Conclusion}
\label{sec:conclusion}
Our work outlines the importance of an application-centric perspective for quantum computing benchmarking. As implementations are typically classical-quantum hybrids, in particular in the NISQ era, various components of the classical-quantum stack need to play together efficiently to achieve high performance. This demands a holistic view on benchmarking procedures.
We present how this approach is realized in the QUARK framework, using modular structures that allow for a simple customization of individual applications. 
Leveraging metrics across the solution pipeline, we investigate trends that help identify paths toward a possible quantum utility.
Six industry-ready use cases from the fields of optimization, machine learning and simulation are analyzed. 
Preliminary results indicate that quantum solutions can reach competitive performance compared to classical implementations also in such industry-relevant scenarios. The benchmarking outcomes help to guide further investigations of possible quantum advantages in the future.

\section*{Acknowledgment}
The project BenchQC is funded by grant BenchQC (DIK-2210-0011//DIK0425/02) of the Bavarian Ministry of Economic Affairs, Regional Development and Energy (StMWi). 
Machine Learning Reply acknowledges contributions of Meghana Lakshmi Narayana and David Gili Fernandez to the hybrid network implementation. 

\bibliographystyle{IEEEtran}
\bibliography{bibliography.bib,bib_settings.bib} 

\end{document}